\documentclass[sigconf]{acmart}

\usepackage{array} %
\usepackage{enumitem}
\usepackage{lineno}
\usepackage{tabularray}
\UseTblrLibrary{booktabs}
\UseTblrLibrary{longtable}
\usepackage{soul}

\AtBeginDocument{%
  }

\setcopyright{acmlicensed}
\copyrightyear{2018}
\acmYear{2018}
\acmDOI{XXXXXXX.XXXXXXX}
\acmConference[Conference acronym 'XX]{Make sure to enter the correct
  conference title from your rights confirmation email}{June 03--05,
  2018}{Woodstock, NY}
\acmISBN{978-1-4503-XXXX-X/2018/06}

\begin{document}
\title[LLM Agent Meets Agentic]{LLM Agent Meets Agentic AI: Can LLM Agents Simulate Customers to Evaluate Agentic-AI-based Shopping Assistants? }

\author{Lu Sun}
\affiliation{%
  \institution{University of California San Diego}
  \city{La Jolla}
  \state{California}
  \country{United States}}
\email{l5sun@ucsd.edu}

\author{Shihan Fu}
\affiliation{%
  \institution{Northeastern University}
  \city{Boston}
  \state{Massachusetts}
  \country{United States}}
\email{fu.shiha@northeastern.edu}

\author{Bingsheng Yao}
\affiliation{%
  \institution{Northeastern University}
  \city{Boston}
  \state{Massachusetts}
  \country{United States}}
\email{b.yao@northeastern.edu}

\author{Yuxuan Lu}
\affiliation{%
  \institution{Northeastern University}
  \city{Boston}
  \state{Massachusetts}
  \country{United States}}
\email{lu.yuxuan@northeastern.edu}

\author{Wenbo Li}
\affiliation{%
  \institution{North Carolina State University}
  \city{Raleigh}
  \state{North Carolina}
  \country{United States}}
\email{wli55@ncsu.edu}

\author{Hansu Gu}
\affiliation{%
  \institution{Independent Researcher}
  \city{Seattle}
  \state{Washington}
  \country{United States}}

\author{Jiri Gesi}
\affiliation{%
  \institution{Independent Researcher}
  \city{Palo Alto}
  \state{California}
  \country{United States}}

\author{Jing Huang}
\affiliation{%
  \institution{Independent Researcher}
  \city{Palo Alto}
  \state{California}
  \country{United States}}

\author{Chen Luo}
\affiliation{%
  \institution{Independent Researcher}
  \city{Palo Alto}
  \state{California}
  \country{United States}}

\author{Dakuo Wang}
\authornote{Corresponding author}
\affiliation{%
  \institution{Northeastern University}
  \city{Boston}
  \state{Massachusetts}
  \country{United States}}
\email{d.wang@northeastern.edu}

\renewcommand{\shortauthors}{Lu et al.}

\begin{abstract}

\textbf{Agentic AI} is emerging, capable of executing tasks through natural language, such as Copilot for coding or Amazon Rufus for shopping. Evaluating these systems is challenging, as their rapid evolution outpaces traditional human evaluation. Researchers have proposed \textbf{LLM Agents} to simulate participants as digital twins, but it remains unclear to what extent a digital twin can represent a specific customer in multi-turn interaction with an agentic AI system. In this paper, we recruited 40 human participants to shop with Amazon Rufus, collected their personas, interaction traces, and UX feedback, and then created digital twins to repeat the task. Pairwise comparison of human and digital-twin traces shows that while agents often explored more diverse choices, their action patterns aligned with humans and yielded similar design feedback. This study is the first to quantify how closely LLM agents can mirror human multi-turn interaction with an agentic AI system, highlighting their potential for scalable evaluation. Our code is open-sourced.~\footnote{https://github.com/neuhai/agentic-ai-evaluation}

\end{abstract}


\begin{teaserfigure}
  \includegraphics[width=1.0\textwidth]{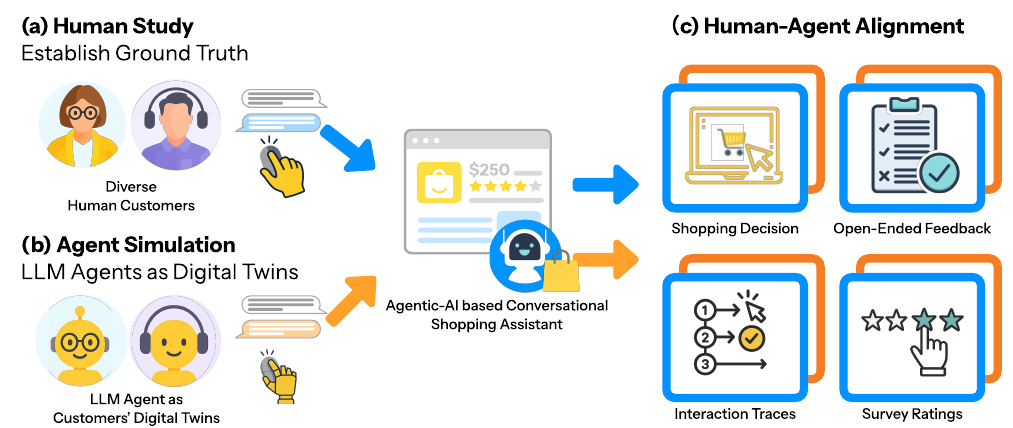}
  \caption{LLM Agent Meets Agentic AI:
(a) We first conduct a human study to establish ground truth about how diverse human users interact with a conversational shopping assistant.
(b) Building on these findings, we then simulate agentic AI agents as digital twins, role-playing the personas of real human participants to reproduce their decision-making behaviors in a controlled browser environment.
(c) Finally, we evaluate human–agent alignment by comparing four aspects—shopping decisions, subjective feedback, interaction traces, and survey ratings—between real users and their simulated counterparts.}
  \Description{Diagram illustrating the workflow of aligning human and agent behavior in a shopping assistant study.
(a) Human Study: Diverse human customers interact with a conversational shopping assistant to establish ground truth.
(b) Agent Simulation: LLM agents act as digital twins of customers, role-playing their decision-making with the same shopping assistant.
(c) Human–Agent Alignment: Researchers compare humans and agents across four aspects, shown as icons—shopping decision, open-ended feedback, interaction traces, and survey ratings.}
  \label{fig:teaser}
\end{teaserfigure}

\maketitle

\section{Introduction}

Agentic AI systems are rapidly emerging across domains, enabling people to accomplish complex tasks through natural language interaction~\cite{firework2023conversational,baier2018conversational}. From GitHub Copilot assisting developers in coding to OpenAI's ChatGPT supporting knowledge work, these systems represent a shift from static tools to proactive and conversational collaborators. 
A particularly impactful application area is e-commerce, where agentic AI transforms traditional search-and-filter shopping into dynamic multi-turn dialogues~\cite{mehta2024rufus}. 
Instead of typing keywords into a search bar, customers can now interact with conversational shopping assistants that adapt to evolving preferences, refine recommendations in real time, and provide personalized guidance throughout the shopping journey. 
This agentic shift is exemplified by systems like Amazon's Rufus~\cite{mehta2024rufus}, Google Shopping's AI mode, and ChatGPT with integrated shopping capabilities~\cite{reuters2025openai}, all of which demonstrate how AI-powered dialogue can reshape the online shopping experience into one that is more personalized, interactive, and engaging~\cite{firework2023conversational,rincon2025shop}.

Despite the rapid deployment and continual iteration of agentic-AI assistants, evaluating their performance and user experience (UX) has not kept pace. New AI features and capabilities can be released on a weekly or even daily basis, yet traditional evaluation methods such as think-aloud, heuristic walkthroughs, A/B testing, or Wizard-of-Oz studies require weeks of planning, recruitment, and analysis \cite{10.1145/3432942, wang2025agenta}. 
These methods have long provided valuable insights~\cite{langevin2021heuristic}, but their slow and resource-intensive nature makes them poorly suited for assessing fast-evolving agentic AI systems in the real world \cite{lu2025uxagent}. 
The problem is compounded by the open-ended and adaptive nature of human-agentic-AI interactions: different users pursue different goals, and the system dynamically tailors its responses, introducing high variability that resists standardized evaluation criteria~\cite{10.1145/3726302.3731955}. 
Taken together, these factors create a widening gap: while agentic-AI systems evolve rapidly, human-centered evaluation struggles to keep up, leaving researchers without scalable tools to capture how these systems shape decision-making, trust, and usability in practice.

To overcome the limitations of traditional UX evaluation methods, recent work has turned to large language model (LLM) agents as scalable evaluators~\cite{zhuge2024agent,zheng2023judging}. The prevailing \textbf{Agent-as-a-Judge} paradigm tasks LLMs with assessing the outputs of other AI models—for example, rating the quality of generated code or verifying factual accuracy, where evaluation criteria are relatively objective and framed in single-turn interactions~\cite{zhuge2024agent,chen2025multiagentasjudgealigningllmagentbasedautomated}. 
Similarly, researchers in the social sciences have used LLM agents to simulate survey respondents or role-play individuals, yielding plausible outcomes such as political attitudes or consumer preferences, yet still confined to single-turn responses. 
While these approaches demonstrate the potential of LLM agents as human \textbf{digital twins} for scalable evaluation, they remain restricted to judging static outputs or isolated prompts. 
Moreover, prior work has largely emphasized algorithmic correctness and agent performance \cite{park2024generative,lu2025promptingneedevaluatingllm}, overlooking user-centered dimensions such as trust, usability, and decision support that are critical for evaluating real-world interactions with agentic AI systems. 
What remains missing is the ability to evaluate \textbf{multi-turn human–AI interactions}, where users dynamically shape the trajectory of a conversation and the system adapts in real time. This gap motivates our central question: \textbf{can LLM agents go beyond judging isolated responses to role-play customers in dynamic multi-turn interaction with agentic AI systems?}

Recent studies have explored LLM-agent simulations in benchmark tasks~\cite{zhu2025automatedriskygamemodeling,park2024generative}. 
For example, the Mind2Web2 benchmark introduced 130 web tasks and constructed task-specific judge agents to automatically assess correctness on search tasks. 
These efforts highlight the potential of agent-based evaluation, but they generally emphasize final outcomes such as negotiated prices or task accuracy, and often operate in constrained environments disconnected from real users' multi-turn behavior. 
In this work, we narrow the focus to online shopping, a domain that is preference-rich, decision-intensive, and inherently conversational in real-world shopping. 
Our study centers on conversational shopping assistants, which can guide customers through discovery, comparison, and decision-making via dialogue. 
A leading example is Amazon Rufus~\cite{mehta2024rufus}, an already-deployed, commercially available system that illustrates how agentic AI can mediate the end-to-end shopping journey via conversational interaction. 
This setting is especially valuable for studying digital twins. It combines wide deployment with diverse users and supports multi-turn interactions that highlight conversation quality. It also produces observable outcomes such as product choices, dialogue trajectories, and UX ratings, which enable user-centered evaluation beyond algorithmic correctness.

To investigate this research question, we designed a two-stage evaluation pipeline that head-to-head compares a human customer's data with their LLM-agent digital twin's data. 
In the first stage, we conducted a large-scale user study in which 40 participants interacted with Amazon Rufus~\footnote{\url{https://www.aboutamazon.com/news/retail/amazon-rufus}} to complete two representative shopping tasks. This produced a rich dataset of multi-turn conversations, product choices, and UX feedback. In the second stage, we instantiated persona-aligned LLM agents via UXAgent~\cite{lu2025uxagent} to role-play as digital twins of the same participants and repeat the tasks. 
This approach enables pairwise comparisons of humans and their digital twins across decision outcomes, interaction behaviors, and evaluation results, providing a foundation for assessing how closely LLM agents can mirror real users in multi-turn interaction with agentic AI systems.

Our analyses reveal that LLM agents can meaningfully approximate human behavior while also exposing important gaps. 
First, agents consistently completed the shopping tasks, matching humans in overall interaction turn counts and final buy-or-not decisions (F1 score of 0.9).
For example, the agent's opening queries showed some alignment with humans' first queries (similarity > 0.4).
This result demonstrates that digital twins can capture the broad structure of human–AI interaction. 
Second, despite these similarities, trajectories soon diverged: sequence-level comparisons showed low overlap (similarity < 0.2), and only about 2\% of agent–human pairs chose the same product.
These differences highlight opportunities to improve how LLMs model human exploration and decision strategies via various fine-tuning techniques. 
Finally, when acting as evaluators, agents produced UX ratings aligned with human judgments on objective dimensions such as query relevance and coherence. 
Yet they tended to rate their own satisfaction more conservatively, while at the same time expressing a stronger preference for interacting with conversational shopping assistants over traditional search. 
Overall, these findings indicate that digital twins can reproduce many functional aspects of shopping interactions while also pointing to concrete directions for training them to better capture the nuance of human decision-making.

Building on these findings, our study demonstrates both the promise and the limitations of using persona-grounded LLM agents as digital twins for evaluating agentic AI systems. Although our analysis is situated in online shopping, the framework and insights extend to other applications where agentic AI engages users in adaptive, multi-turn interactions. More broadly, this work contributes to the methodological toolkit of human–AI interaction by combining the breadth of agent-based simulation with the depth of human-centered evaluation.

In sum, our contributions are:
\begin{itemize}[noitemsep, topsep=0pt]
\item The first large-scale human study of multi-turn interactions with an agentic-AI–powered conversational shopping assistant (Amazon Rufus), capturing buy-or-not outcomes, interaction traces, and UX ratings.
\item The first simulation framework that instantiates persona-grounded LLM agents as digital twins, enabling direct pairwise comparison with real customers across tasks, behaviors, and evaluations.
\item Empirical insights into where agents align with or diverge from humans, providing the first evidence of both their potential as scalable evaluators and their limitations in capturing human-like reasoning and experience.
\end{itemize}

\section{Related Work}

\subsection{From Traditional Online Shopping to Agentic-AI-based Shopping Assistants}

Agentic AI systems are rapidly emerging and reshaping how users engage in online activities across multiple domains~\cite{acharya2025agentic, shavit2023practices, sapkota2025ai, murugesan2025rise}. 
For example, GitHub Copilot~\cite{github_copilot} is transforming software development through conversational code generation~\cite{wermelinger2023using, zhang2023practices}, while Google Gemini~\cite{google_gemini} is redefining search by enabling natural language–based exploration~\cite{wazzan2024comparing, spatharioti2025effects, mo2025conversational}. 
These systems are capable of executing complex tasks through iterative interactions in natural language~\cite{Yun2025uSER, Vaithilingam2022}, lowering barriers to access and broadening participation~\cite{Schneiders2025, Zamfirescu2023}.

Among these, Conversational Shopping Assistants (CSAs) have quickly become one of the most visible applications. 
In early 2025, several prominent companies—including Amazon~\cite{mehta2024rufus}, Google~\cite{marinlopez2025google,rincon2025shop}, and OpenAI~\cite{reuters2025openai}—introduced their own versions of CSAs, which rapidly gained widespread attention and adoption.
Traditionally, online shopping required users to rely on search boxes and filters, manually navigating across multiple tabs to browse, compare, and ultimately select products~\cite{haubl2000consumer, rowley2000product, hong2004designing, moser2017no}. 
By contrast, CSAs have begun to transform this process~\cite{firework2023conversational,baier2018conversational}. 
Instead of keyword searches and static filters, users can now engage in multi-turn, natural language conversations with LLM-based CSAs to explore product options, refine their preferences, and receive tailored recommendations~\cite{Yitian2025}.

\begin{figure}[htb]
    \centering
    \includegraphics[width=.8\linewidth]{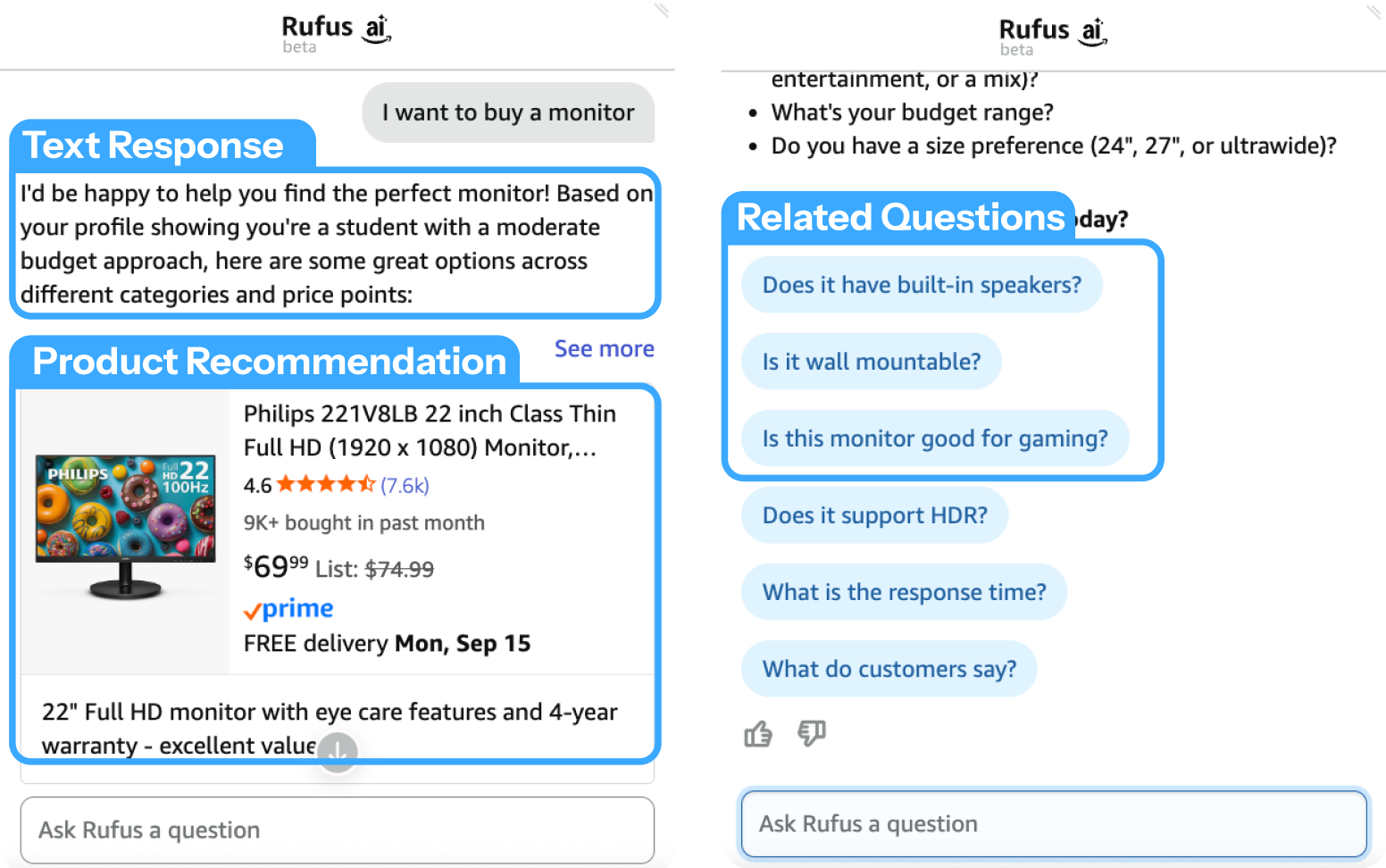}
    \caption{Amazon Rufus Conversational Shopping Assistant. The user can send in a natural language query to Rufus. Rufus can respond with text responses, product recommendation cards, and related question suggestions.}
    \Description{Screenshot of Amazon Rufus conversational shopping assistant. On the left, a user query “I want to buy a monitor” receives a text response with shopping guidance, followed by a product recommendation card showing a Philips 22-inch monitor with price, rating, and delivery details. On the right, Rufus suggests related questions such as “Does it have built-in speakers?”, “Is it wall mountable?”, and “Is this monitor good for gaming?”.}
    \label{fig:rufus}
\end{figure}

A typical Conversational Shopping Assistant (CSA) system like Rufus (Fig.~\ref{fig:rufus}) is seamlessly integrated into the shopping website interface
Users can open the CSA through a pop-up window available from the navigation bar on any page, making the assistant easily accessible without interrupting the normal shopping flow. 
During browsing or product search, users can engage with the CSA at any time to request personalized product recommendations, seek clarifications, or compare alternatives.
CSA supports natural language queries (e.g., ``I want to buy a monitor'') and responds in a multimodal format that includes explanatory text responses, product recommendation cards with images and pricing, as well as related question suggestions that guide further exploration. This interaction loop allows users to iteratively refine their preferences, verify product attributes, and make more informed purchase decisions, thereby augmenting the traditional search-and-filter paradigm with a conversational and context-aware shopping experience.

\subsection{Challenges in Evaluating Agentic AIs}
Evaluating the usability of agentic AI systems, especially those involving conversational assistants, presents unique methodological challenges~\cite{luger2016, clark2019makes, brandtzaeg2018chatbots}.
First, agentic AI can exhibit highly adaptive behaviors~\cite{wu2022AIchains}. 
Even when faced with the same task prompt, different users may choose distinct strategies, phrasing, or levels of detail in their inputs~\cite{Weisz2024}. 
As a result, the same task can unfold differently across users and even across sessions with the same user~\cite{Zamfirescu2023}.
This variability creates challenges for evaluation, as standardized usability measures—such as task completion time or error rates—may not capture the full spectrum of user behaviors and contexts~\cite{kocielnik2019will, zheng2025}. 
For example, two users might reach the same outcome through very different interaction paths, making it difficult to determine whether one experience is objectively “better” than the other.
Second, users' perceptions of response quality and task support remain inherently subjective~\cite{Reinecke2011, duan2025systematic}. 
These perceptions are shaped not only by the correctness or completeness of system outputs, but also by individual goals, prior knowledge, cognitive styles, and situational needs~\cite{zheng2025enhancing}.
What one user considers a helpful explanation, another may view as excessive or confusing. Moreover, perceptions can shift dynamically during interaction: a user who initially values speed may later prioritize detail and transparency once they gain confidence in the system~\cite{xiao2007role, koh2025understanding}.
Such subjective and evolving evaluations further complicate the assessment of system effectiveness, as measures of satisfaction, trust, or usefulness may vary widely both across and within users~\cite{khurana2024and}.

Traditional usability testing approaches often rely on human-centered methods that directly involve researchers, experts, or participants~\cite{gulliksen2003key}. 
These methods are designed to uncover usability issues, evaluate user experience, and ensure that systems align with human needs~\cite{sauro2016quantifying, hartson2012ux}.
For example, expert heuristics~\cite{langevin2021heuristic, nielsen1994usability} involve evaluators applying established principles to identify usability flaws early in the design process, but their insights can be limited by the evaluators' expertise and may overlook domain-specific challenges~\cite{ammenwerth2003evaluation, andre2001user}. 
Controlled user studies~\cite{sharp2007interaction, lazar2017research} provide empirical data through carefully designed tasks and measures, yet they require significant time, participant recruitment, and researcher effort to conduct, making them costly and difficult to repeat at scale~\cite{kjeldskov2003review}. 
Wizard-of-Oz methods~\cite{dahlback1993wizard, kuang2024enhancing} allow researchers to prototype interactive systems by simulating system responses through human operators, offering rich insights into user expectations and interaction patterns, but they are labor-intensive and difficult to sustain beyond small studies~\cite{kelley1984iterative, dow2005wizard}.

While these methods have been foundational in HCI and continue to offer valuable insights, they are resource-intensive and may not scale well for rapidly evolving agentic AI systems. 
This limitation is particularly evident in the context of LLM-based conversational assistants, where system behavior is highly dynamic, responses are non-deterministic, and user interactions can vary significantly across sessions~\cite{kuang2023collaboration, chen2025toward}. 
As a result, traditional approaches struggle to keep pace with the scale, speed, and variability of modern AI systems, highlighting the need for complementary methods of evaluation.

\subsection{LLM Agents as Proxies for Human Participants in UX Evaluation}

Since involving human participants in the evaluation of AI-powered systems requires substantial time and effort, researchers have increasingly explored using agents as evaluators of other systems.
LLM-as-a-Judge~\cite{zheng2023judging} is one such approach, where an LLM is used as a judge to compare the outputs of different models on the same task. In benchmarks like Chatbot Arena~\cite{zheng2023judging} and MT-Bench~\cite{zheng2023mtbench}, this method is commonly used to determine which response is better while significantly reducing the cost of human evaluation.
Building on this idea, the Agent-as-a-Judge framework~\cite{zhuge2024agent} extends LLM-as-a-Judge by equipping the judge with agentic capabilities such as autonomous planning, information retrieval, step-by-step execution, and tracking intermediate artifacts. This enables the judge to function as an agent and evaluate the entire task-solving process rather than just the final output, demonstrating that prompt-driven LLM agents can perform complex evaluation tasks traditionally handled by human experts.
Meanwhile, broader evaluation suites such as AgentBench\cite{liu2023agentbench}, MLAgentBench\cite{huang2023mlagentbench}, and Mind2Web2\cite{gou2025mind2web} provide standardized tasks for measuring LLM agents' capabilities.
For example, Mind2Web2\cite{gou2025mind2web} evaluates how well AI agents handle realistic, long-horizon, and dynamic web search tasks. It adopts an Agent-as-a-Judge framework, where a judge agent verifies whether each answer satisfies all task requirements and includes reliable citations. This setup allows researchers to systematically evaluate agentic search systems on complex real-world tasks.
More specifically, in the context of online shopping, ~\citet{zhu2025automatedriskygamemodeling} proposed an Agent-to-Agent simulation framework to systematically study how LLM-based agents negotiate and make decisions on behalf of human consumers or merchants.
In this framework, LLM agents simulate buyers and sellers in consumer markets, engaging in multi-turn dialogues to negotiate prices and complete transactions. This allows the researchers to evaluate each agent's decision-making strategies and negotiation capabilities.
However, such evaluations mainly focus on the final negotiation outcomes (e.g., agreed prices or profit margins) rather than the fine-grained interaction dynamics that typically occur when users interact with conversational shopping assistants.
Yet, this line of research centers on technical performance, paying little attention to the diverse needs and intentions that shape how different user groups engage with AI-powered conversational systems.

In response, researchers have begun examining how LLM agents can generate actions that resemble those of diverse human users.~\cite{wangSurveyLargeLanguage2024, xie2024can}. 
For example, LLM agents have been used to emulate a community of 25 residents in a virtual village~\cite{parkGenerativeAgentsInteractive2023}, to replicate participants in social science studies~\cite{parkGenerativeAgentSimulations2024,schmidgallAgentClinicMultimodalAgent2024,leeApplicationsGPTPolitical,gurcanLLMAugmentedAgentBasedModelling2024}, to act as patients and clinicians in hospital contexts~\cite{liAgentHospitalSimulacrum2024}, and to take on the role of software developers in company settings~\cite{qianChatDevCommunicativeAgents2024}.
\begin{figure*}[t]
    \centering
    \includegraphics[width=0.75\linewidth]{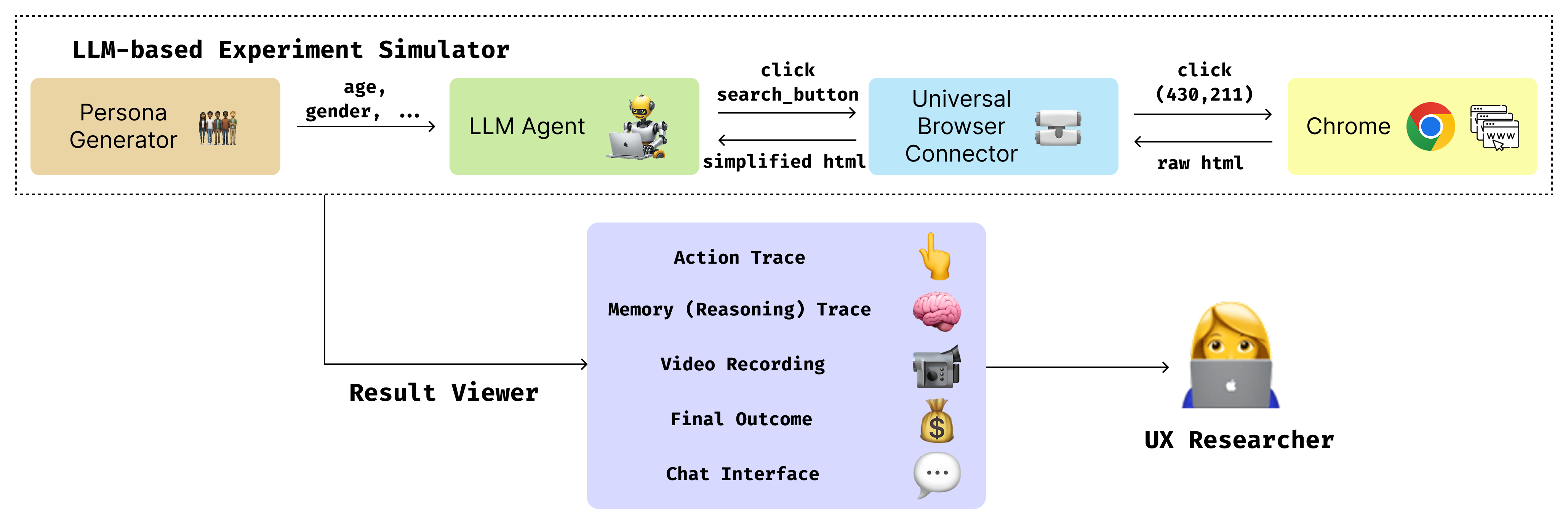}
    \caption{System Architecture of UXAgent \cite{lu2025uxagent}}
    \Description{Diagram of the UXAgent system architecture. At the top, the LLM-based experiment simulator includes a persona generator feeding demographic traits (e.g., age, gender) to an LLM agent. The agent issues actions such as “click search button” through a universal browser connector, which translates them into raw HTML interactions with Chrome. At the bottom, outputs from the simulator (action trace, memory trace, video recording, final outcomes, and chat interface) are sent to a result viewer, which UX researchers use to analyze agent behavior.}
    \label{fig:uxagent}
\end{figure*}

Building on these advances, the use of LLM agents for the evaluation of user experience also shows promise in various domains,~\cite{wang2023user,xiang2024simuser,wang2025agenta, ren2014agent}, including graphical user interface testing~\cite{eskonenAutomatingGUITesting2020}, game environments~\cite{stahlkeArtificialPlayfulnessTool2019,fernandesAgentsAutomatedUser2021}, and accessibility evaluation~\cite{taebAXNavReplayingAccessibility2024,zhong2025screenaudit}.
For instance, SimUser~\cite{xiang2024simuser} introduced a dual-agent framework in which one agent simulated user behaviors and another emulated the mobile app interface. While this approach showed promise for generating behavioral data automatically, it was limited by its reliance on simulated environments.
As shown in Fig. \ref{fig:uxagent}, UXAgent~\cite{lu2025uxagent} introduced a framework where LLM agents simulate users performing tasks on web pages, enabling automated heuristic evaluations. 
The persona generator generates diverse agent personas at scale, and the LLM Agent interacts with Web Browser through Web Browser Connector and produces usability data for the UX Researcher to analysis.
Building on this, AgentA/B~\cite{wang2025agenta} applies simulated users in A/B testing contexts, demonstrating that LLM agents can detect usability improvements across interface variants in a reproducible and scalable manner. 

In summary, existing approaches still lack grounding in empirical, ground-truth data derived from diverse real human participants. 
To address this gap, we first conducted a human-subject study to evaluate their user experiences with existing conversational shopping assistants. And conduct an agent simulation based on a real human user study.

\section{Method}

Our study consisted of two stages: (1) a~\textbf{\textit{human study}} where we invited human participants to completed two shopping tasks with conversational shopping assistants and conducted user evaluations on its usability, engagement, trust, etc and (2) an~\textbf{~\textit{role play agent simulation}} that created a digital twin agent of each participant and performed the same task procedure and evaluation procedure. 

In Stage 1, we recruited 40 participants to complete two shopping tasks with conversational shopping assistants. We collected demographic and background information to design realistic personas, and logged 80 shopping sessions. These sessions captured interaction traces and usability evaluations (e.g., satisfaction, task success, and perceived helpfulness), establishing a baseline for comparison with agent simulations.

In Stage 2,  we scaled this evaluation through an Agent-as-a-Judge simulation. Using personas obtained from the human study, LLM-based agents role-played as digital twins for each participant, where they completed the same tasks and provided evaluations. This allowed us to assess how closely simulated agents can reproduce human behaviors and UX judgments.

Specifically, we aim at answering three research questions (RQ) in our two-stage study:
\begin{itemize}[leftmargin=0pt, label={}, nosep]
    \item \textbf{RQ1.} How do human customers interact with and evaluate conversational shopping assistants (CSAs)?
    \item \textbf{RQ2.} To what extent can LLM agents role-play as customers when performing shopping tasks?
    \item \textbf{RQ3.} How closely do agent-based customer simulations align with human behaviors in task outcomes, interaction patterns, and user experience evaluations?
\end{itemize}

\begin{figure*}[htb]
    \centering
    \includegraphics[width=1.0\linewidth]{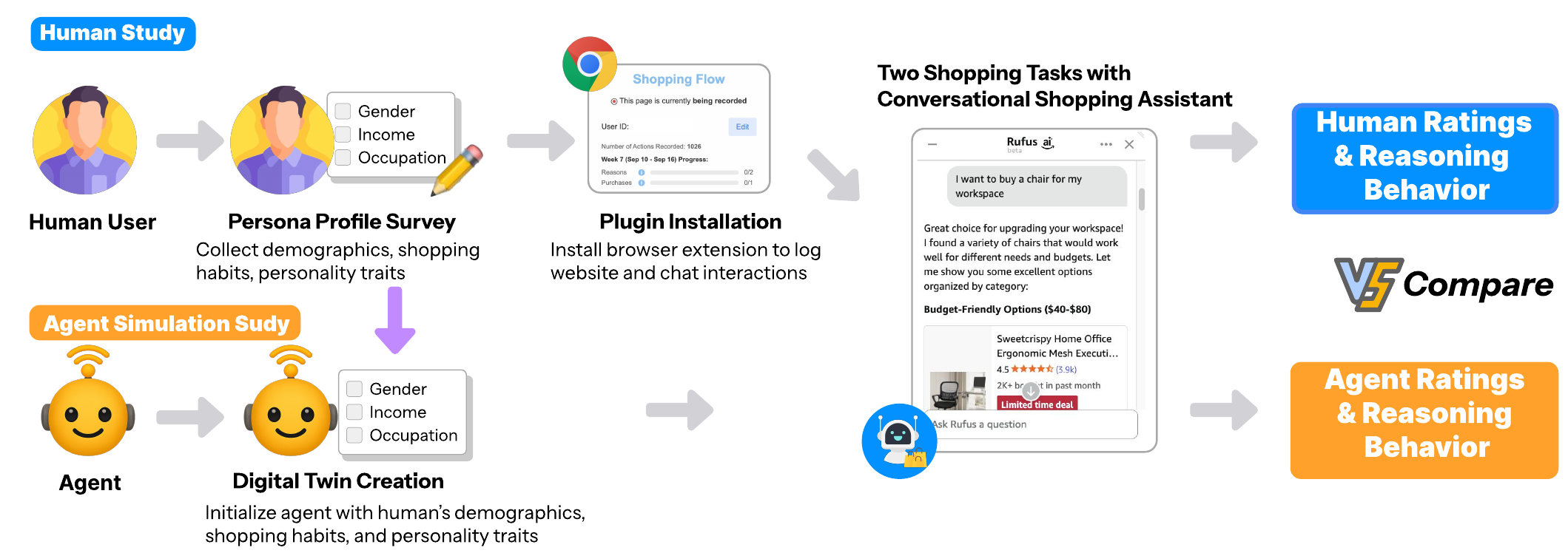}
    \caption{Study Procedure Overview:
We first collected human participants' demographics, shopping habits, and personality traits, then logged their interactions while they performed shopping tasks with a CSA. Next, we injected the same personas into LLM-based agents to complete the same tasks, and compared their ratings and reasoning behaviors to assess human–agent alignment.}
    \label{fig:procedure}
    \Description{Diagram of study procedure. At the top (Human Study), a human user completes a profile survey on demographics, shopping habits, and personality traits, installs a browser plugin, and performs two shopping tasks with a conversational shopping assistant. Their ratings and reasoning behavior are collected. At the bottom (Agent Simulation Study), an agent is initialized as a digital twin with the same demographic and personality traits, then performs the same shopping tasks. Human and agent ratings and reasoning behaviors are compared to assess alignment.}
\end{figure*}

\subsection{Stage 1 Human Study: Establish Ground Truth}~\label{sec:humanstudy}

\subsubsection{\textbf{Participants}}
We recruited 50 participants from Prolific, aiming for a diverse U.S. sample across age, gender, race, region, education, and income level. Of these, 40 participants completed all tasks and surveys, and their data were included in our analysis. As part of a pre-screening survey, potential participants were asked to confirm that they had used an online shopping platform before, such as Amazon or Google Shopping. All studies are conducted remotely, where participants follow an instruction page that is designed to guide them through the 4-stage procedure. All participants are compensated by \$15.5/h. The study protocol received ethical approval from the university's Institutional Review Board.

\subsubsection{\textbf{Study Context - Amazon Rufus}}
We conducted a user study in which participants interacted with and evaluated an agentic-AI-based conversational shopping assistant—Amazon Rufus~\footnote{\url{https://www.aboutamazon.com/news/retail/amazon-rufus}}
. Amazon Rufus is designed to support product discovery and decision-making through natural language dialogue. It allows users to describe their needs in free-form text, ask follow-up questions, and receive tailored product recommendations, as shown in Fig~\ref{fig:rufus}. Beyond simple keyword search, Rufus can summarize product features, compare alternatives, and iteratively refine suggestions across multiple conversational turns. Integrated directly into the Amazon shopping platform, it enables seamless transitions from exploration to purchase within the same web shopping interface. We selected Amazon Rufus for this study because it is one of the most widely deployed conversational shopping assistants in commercial use today and offers a highly representative setting for studying user behaviors, expectations, and perceptions in shopping scenarios.

\subsubsection{\textbf{Procedure}}
The average study duration is 37.5 minutes. The overall procedure is illustrated in Figure~\ref{fig:procedure}. 

\paragraph{Step 1: Demographic Survey}
Upon entering the study, participants first completed a 15-minute structured online survey designed to elicit detailed persona profiles. The survey included items adapted from prior work shown to correlate with consumer behavior~\cite{hou2021mobile,wang2025opera}. It consisted of three main sections: demographic information, shopping preferences, and personality traits. Demographics covered age, gender, ethnicity and race, education, occupation, household income, and residence~\cite{hou2021mobile}. Shopping preferences covered topics such as online shopping frequency~\cite{shao2014impact}, membership status, shopping habits, seasonality~\cite{nizam2018interactive}, trust in advertising~\cite{shao2014impact}, engagement with product reviews, influence of delivery options~\cite{bauboniene2015commerce}, and an adapted eight-item Consumer Styles Inventory (CSI)~\cite{nayeem2022revisiting}. Personality traits were measured using the Big Five Inventory~\cite{goldberg1992development} and a self-reported Myers-Briggs Type Indicator (MBTI)~\cite{myers2003mbti}. To contextualize each persona's familiarity with the shopping tasks, we also collected self-reported experience levels across the four product categories: monitors, chairs, outfits, and jackets. Finally, to further enrich the persona profiles, open-ended questions invited participants to describe their daily routines, shopping habits, and consumer identity\cite{park2024generative}. Full survey items are in the Appendix.

\paragraph{Step 2: Plugin Install}
Participants were asked to install a custom Chrome extension designed to capture their interactions with both the webpage and the conversational shopping assistants (CSAs). To support detailed interaction logging, we adapted the ShoppingFlow framework~\cite{wang2025opera} extended the browser plugin that automatically records user behaviors and contextual signals during shopping sessions not only on the Amazon search page, but also with the Amazon Rufus Chatbot. 
The extension tracks granular actions such as text inputs, clicks, scrolls while interacting with Rufus, and on Amazon.com, as well as browser-level events including navigation, tab switching, and page reloads. 
Specifically, we focused on users' behaviors with Rufus, including typing and responding with Rufus chatbot, clicking on related questions, and clicking on recommended product, as shown in Figure~\ref{fig:rufus}. To better understand users' decision-making processes, the extension also featured an in-situ reflection mechanism:  when participants click on related questions and recommended products, participants were prompted with a lightweight pop-up asking them to briefly explain their reasoning behind specific interactions, as shown in table~\ref{tab: pop_up_question}.

\begin{table}[t]
    \centering
    \caption{Examples of in-situ pop-up questions prompting participants to explain their interaction choices.}
    \begin{booktabs}{
    colspec={X[1]X[2]},
    row{1}={c,font=\bfseries}
    }
    \toprule
        Participant Action & Pop-up Questions\\ \midrule
        \textit{click} on related questions & Why do you want to ask rufus chatbot on {user questions}? \\ 
        \textit{click} on recommended products  & We noticed that you click on {product}. Why did you do this? \\ 
        \textit{type} in the chat input & You typed in the chat input. What do you want to ask for help?\\ 
        \bottomrule
    \end{booktabs}
    \label{tab: pop_up_question}
\end{table}

\paragraph{Step 3: Conduct Two Hypothetical Shopping Tasks}
Participants then completed shopping sessions using the Amazon Rufus chatbot across four tasks designed to represent both utilitarian (buying a monitor or a chair for home use) and hedonic (choosing an outfit for a summer wedding or a jacket for a group hike) shopping motivations (see Table~\ref{tab:task description}). To control for task and product domain effects, participants were randomly assigned to one of two groups. One group completed tasks involving the purchase of a monitor and selection of a wedding outfit, while the other group completed tasks focused on purchasing a chair and selecting a hiking jacket. This grouping ensured coverage across distinct shopping goals and enabled comparison across different product categories.

\begin{table}[t]
    \centering
    \small
    \caption{Participant shopping tasks of two groups. Each task specifies a category, a concrete goal, and detailed instructions to be completed. Each group completed an utilitarian task and a hedonic task. }
    \label{tab:task description}
    \begin{booktabs}{
        colspec={X[0.8]X[1.6]X[3]},
    cells={c,m},
    width=\linewidth,
    column{3}={l},
    row{1}={c,font=\bfseries},
    }
    \toprule
     \textbf{Category} & \textbf{Task} & \textbf{Description} \\
    \midrule
    Utilitarian task & Find a monitor for home use &
    Use Rufus Chatbot to pick a \$800 chair that is ergonomic for long work hours for your home workspace and add it to your shopping cart.  \\
    Hedonic task & Find an outfit for a summer wedding &
    
    You've been invited to a hiking event with friends. Use Rufus Chatbot to to explore hiking jackets within a \$200 budget that is waterproof.  If you find a desired product, add it to your shopping cart to show your intent to purchase.
     \\
    \midrule
    Utilitarian task &Find a chair for home use &
    
   Use Rufus Chatbot pick a \$400 monitor with the highest possible resolution for your home workspace. 
     \\
    
     Hedonic task & Find a jacket for a group hike &
   
    You've been invited to a summer wedding with a green theme. Use Rufus to explore your outfit options within a \$200 budget. If you find a desired product, add it to your shopping cart to show your intent to purchase.
     \\
    \bottomrule
    \end{booktabs}
\end{table}

\paragraph{Step 4: Finish UX Evaluation Survey}
After the study session, participants filled out a survey assessing satisfaction, perceived utility, and overall experience with the assistant. After completing each task, participants filled out a user evaluation survey assessing their satisfaction, perceived usability, engagement, cognitive effort, trust, and overall experience with the shopping assistant. Appendix lists the full user evaluation survey items.

\subsubsection{\textbf{Plugin System Development}}
To support detailed interaction logging, we adapted the ShoppingFlow framework~\cite{wang2025opera} and developed a custom browser plugin that automatically records user behaviors and contextual signals during shopping sessions with Amazon Rufus. The plugin consists of two main components: a Content Script and a Background Script. The Content Script operates within the foreground of the active webpage, capturing fine-grained user interactions such as clicks, text inputs, scrolls, and DOM content changes. Each interaction is time-stamped and annotated with contextual metadata, including CSS selectors, semantic labels (e.g., \texttt{search\_result.product\_title}), and DOM attributes to enable precise mapping of actions to interface elements. The Background Script complements this by monitoring higher-level browser events such as page navigations, tab switches, and reloads. Together, this dual-script architecture ensures comprehensive coverage of both micro-level interactions and macro-level session dynamics, while maintaining efficient performance. To better understand participants' decision-making processes, the plugin also includes an in-situ reflection mechanism. When users interact with certain UI elements—such as clicking on recommended products or related chatbot questions—the system triggers lightweight pop-up prompts asking them to briefly explain their rationale for the action.

All captured data—including user action traces, DOM snapshots, simplified HTML, and rationale responses—were stored and uploaded in real time to a secure Amazon S3 bucket whenever a user-triggered event occurred (e.g., clicking a product, typing a query, navigating to a new page).
To safeguard user privacy, the plugin was explicitly configured to exclude personally identifiable information (PII) by skipping sensitive pages such as login, profile, and checkout flows. Additionally, we implemented a rule-based automated script that detects and masks any residual PII—such as usernames embedded in navigation bars, zip codes, addresses, or payment details. This instrumentation enabled the construction of a rich, multimodal dataset comprising timestamped user actions, contextual web observations, rationale annotations, and detailed persona metadata.

\subsubsection{\textbf{Measurements}}
To assess both human and agent-based evaluations of conversational shopping assistants, we developed a three-part evaluation framework covering task outcome, interaction quality, and user experience. These metrics were designed to capture not only whether users completed the shopping task successfully, but also how they interacted with the assistant and how they perceived the experience. Human evaluation data were collected through post-task surveys and system-logged behavioral traces. Simulated agents were evaluated using analogous interaction logs and inferred signals. These items captured key dimensions of user experience, including task success, user satisfaction, and user perceptions of the interaction—such as product information accuracy, usability, efficiency, conversational quality, trust, and cognitive load.

\paragraph{Shopping Task Outcomes}
We measured task success based on a combination of user self-reports and final product selections. After each shopping session, participants confirmed whether they used the assigned assistant (i.e., Amazon Rufus) and indicated whether they were able to find a product that matched the given shopping goal. Participants also rated their satisfaction with the final outcome on a 5-point Likert scale, providing a subjective measure of task success. 

\paragraph{User Interaction Data}
Interaction quality was assessed using a combination of automatically logged behavioral data from the Chrome extension and self-reported user feedback. From the interaction logs, we analyzed the number of clarification queries made during each session, reflecting how actively users engaged in refining their requests and how well the assistant supported iterative dialogue. We also examined the depth and specificity of user queries, particularly when users asked about nuanced product features, comparisons, or trade-offs—indicating the extent to which the assistant enabled informed decision-making. 

\paragraph{User Experience Survey}
Participants completed a short post-task survey rating their experience with the CSAs. We captured their evaluations on usability, engagement, satisfaction, trust, and cognitive effort.  Usability was captured using adapted items from the System Usability Scale (SUS)~\cite{brooke1996sus}, such as ``It was easy to interact with Rufus'' and ``I found the interaction enjoyable.'' Satisfaction and intention to reuse were measured using items like ``I would love to use Rufus to shop in the future'' and ``I will recommend others to use Rufus to shop.'' Perceived helpfulness was assessed through statements such as ``Rufus helped me narrow down product options.'' Conversational engagement was also measured with the statement ``The conversation with Rufus felt engaging.''

To complement behavioral data, participants also self-reported their engagement levels and cognitive effort~\cite{davis1989perceived,hart1988development}. Example statements included ``I found the interaction mentally demanding'' and ``I had to put in a lot of effort to use Rufus effectively,'' both adapted from the NASA Task Load Index (NASA-TLX)\cite{hart1988development} and the Technology Acceptance Model (TAM)\cite{davis1989perceived}.

In addition, participants evaluated the assistant's information accuracy and performance~\cite{balakrishnan2024conversational}.  These included items such as ``Rufus provided accurate and up-to-date product information,'' ``Rufus's responses were logical and coherent,'' and ``Rufus described product features in a way that matched official information,'' which collectively measured perceived information accuracy, coherence, and factual consistency. 

We also included measures of perceived trustworthiness and reliance, which are essential in assessing user confidence in AI recommendations. Participants rated statements such as ``I trust the responses provided by Rufus,'' ``I would rely on Rufus without double-checking its responses,'' and a reverse-coded item: ``I was concerned that Rufus may present biased or sponsored recommendations.'' These items reflect emerging concerns around AI bias, transparency, and user trust in conversational agents~\cite{balakrishnan2024conversational}.

At the end of the survey, participants were asked an open-ended question: ``What do you like about Rufus? What do you dislike?'' This allowed us to capture qualitative insights into user-perceived strengths, weaknesses, and unmet expectations that may not be captured by structured response scales. Together, these measures provide a multidimensional view of user experience across usability, utility, trust, cognitive effort, and emotional response, enabling both quantitative comparison and in-depth qualitative interpretation.

\subsubsection{\textbf{Data Analysis}}

To establish a human benchmark, we analyzed data from 40 participants who completed four shopping tasks (monitor, chair, summer outfit, hiking jacket), spanning both utilitarian and hedonic goals. Task success was binary-coded based on whether the selected product met task-specific constraints (e.g., staying within budget, suitability for the occasion). Post-task user experience (UX) ratings were collected on five dimensions—satisfaction, trust, usability, helpfulness, and cognitive load—using 5-point Likert scales (1 = strongly disagree, 5 = strongly agree).

We computed descriptive statistics and conducted paired and independent-sample \textit{t}-tests to compare task outcomes and UX ratings across task types (utilitarian vs. hedonic). Interaction behavior metrics—including total turn count, number of clarification queries, and average message length—were extracted from logged sessions and analyzed.

\subsection{Stage 2 LLM Agent Simulation: Role-Playing as Digital Twins}

In Stage 2, we extended the formative human study by scaling the evaluation using an Agent-as-a-Judge simulation approach. Building on real participant data, we instantiated LLM-based agents conditioned on participant personas, prompting them to role-play as diverse online shoppers. These digital twin agents completed the same shopping tasks as human participants and generated UX evaluations, enabling large-scale, repeatable assessments of conversational shopping assistants—grounded in real-world behavioral patterns.

To simulate human experiences and evaluate AI-powered shopping interactions at scale, we employed the UXAgent framework~\cite{lu2025uxagent}. UXAgent is a persona-driven LLM agent architecture designed to model user behavior through realistic task execution and reflective UX reasoning. In our study, UXAgent was used to simulate \textbf{``digital twins''} of real participants from the human study, enabling direct comparisons between human and agent performance and perception under identical task conditions. We selected UXAgent because of its explicit design for planning and reflection during UX evaluations, which made it well suited to our study goals.

\subsubsection{\textbf{Transform from Real Participants to Persona}}
We grounded our agent simulations in the behavioral, contextual, and evaluative patterns observed in the human study (Section~\ref{sec:humanstudy}). Using demographic profiles, shopping goals, interaction traces, and post-task feedback from 50 participants, we constructed a one-to-one mapping between each human user and their simulated digital twin. Each agent was instantiated with a benchmark persona that combined demographic and contextual attributes—such as age group, gender, shopping frequency, tech familiarity, and budget sensitivity—with shopping styles and priorities inferred from observed behavior, including emphasis on price, brand loyalty, reliance on product comparisons, and the use of clarification strategies. To enrich the persona beyond static attributes, we also incorporated participants' self-descriptions and daily routines, collected through open-ended survey questions. The resulting persona descriptions served as natural language prompts to guide agent behavior, ensuring that each ``digital twin'' reflected not only the participant's demographic background and shopping logic, but also their lived experience and decision-making tendency.

\subsubsection{\textbf{Procedure}}

The simulation followed four core stages:

\paragraph{Persona Initialization.}
Each agent was primed with a natural language persona prompt summarizing the participant's background, shopping motivations, and task-specific constraints. For example:  
\textit{``You are a 34-year-old professional who prefers eco-friendly products under \$400. You are shopping for a monitor for your home office. You care about display quality and customer reviews.''}  
This setup enabled the agent to behave consistently with the user's profile throughout the task.

\paragraph{Plan Generation and Task Execution.}
The agent then begins task execution in an iterative loop of planning and acting. In each iteration, it generates or revises a structured, step-by-step plan to complete the assigned shopping task (e.g., product search, comparison, clarification, selection), grounded in the persona's goals and constraints. The agent then executes this plan via UXAgent's \textit{Universal Web Connector}~\cite{lu2025uxagent}, which enables operation in real-world web environments (including Amazon Rufus) using simplified HTML observations and task-agnostic action primitives.

The Web Connector filters raw HTML into a semantic representation that preserves meaningful interface elements such as product titles, chatbot responses, buttons, and related questions, while discarding visual clutter (CSS, JavaScript). Based on this abstraction, the agent selects from a fixed set of atomic actions: \texttt{click}, \texttt{type}, \texttt{type\_and\_submit}, \texttt{clear}, \texttt{back}, and \texttt{terminate}. These actions are executed through a backend to simulate realistic web interactions.

Each action prediction was made using the Claude 3.7 Sonnet model, selected for its strong reasoning and instruction-following capabilities. The model operated under a low-temperature decoding setting (\texttt{temperature=0.2}) to prioritize consistency and reproducibility of behavior. For each step, the model received the persona context, current task goal, interaction history, and simplified DOM snapshot as input, and predicted the next best action. This dynamic reasoning loop—plan, act, observe, revise—enabled agents to adapt to unexpected responses, revise queries, or switch strategies mid-task, mimicking realistic user behavior.

\paragraph{Post UX Evaluation.}
Upon completing each shopping task, the agent was prompted to evaluate the assistant's performance using their interaction tracing. We asked agents to answer the same survey questions as users.  The agent answers the open-ended questions and rates the Likert-style ratings (1--5) across the same dimensions used in the human study: task success, usability, conversational quality, trust, and cognitive load. These role-aligned self-assessments leveraged UXAgent's prompting strategy to ensure consistent and interpretable judgments.

\paragraph{Interaction and Session Logging.}
All interaction data, including the agent's queries, Rufus's responses, selected actions, clicked elements, and internal planning steps, were logged in detail. We also stored the action traces and final product selections. This enabled fine-grained comparison with human users.

This end-to-end simulation pipeline enabled controlled, persona-consistent evaluation of conversational shopping assistants at scale. By integrating UXAgent's reasoning architecture with its Universal Web Connector for real-world web execution, we were able to simulate human-like shopping behavior on Amazon Rufus and evaluate agentic alignment with real user goals, preferences, and perceptions.

\subsubsection{\textbf{Measures and Analysis}}
To evaluate how well persona-grounded LLM agents align with human users, we conducted a multi-level comparative analysis grounded in the same dimensions assessed in our human study. Simulated agent behaviors were evaluated using both structured outputs (e.g., UX ratings generated at the end of each task) and detailed interaction logs that are automatically captured during the simulation. The analysis was designed to support both qualitative comparison and statistical testing of alignment between humans and their digital twin agents.

\paragraph{Persona-Level Alignment.}
We assessed whether the agent's final product choices were consistent with the preferences encoded in its assigned persona. Each agent was grounded in a real participant's demographic profile, shopping goals, and stated priorities. We compared final product selections between the agent and the corresponding human participant to measure choice consistency.

\paragraph{Task Outcome Alignment.}
To evaluate task success, we calculated the success rate of purchasing products. Then, we computed the F1 score on their decision on purchase or not.

\paragraph{Interaction Behavior Alignment.}
We examined behavioral measures extracted from simulation logs, including the number of dialogue turns, number of recommendation products, and number of related questions. These metrics were directly comparable to human interaction traces. We used Welch's t-test to compare human with their paired agent. We also compared the semantic of messages including the length of messages and cosine similarity of messages. To compare the action trajectory between agent and human, we further calculated the Levenshtein distance of their action sequences. This analysis allowed us to capture both structural alignment and strategic divergence.

\paragraph{UX Evaluation Alignment.}
At the end of each session, agents generated structured UX evaluations using prompts aligned with the human post-task survey. Ratings were captured across the same five dimensions: task satisfaction, perceived helpfulness, usability, conversational quality, and cognitive load. While human ratings were self-reported, agent ratings were inferred via reflective prompting based on the agent's persona and full action trace. To compare these ratings, we used Welch's t-test on the Likert-scale responses, quantifying overestimation tendencies (e.g., agents consistently rating satisfaction higher).

\section{Result}
We analyzed 40 participants' data across 80 human shopping sessions and their digital twins' data across 80 agent-simulated sessions to answer research questions.

\subsection{RQ1: How do human customers interact with and evaluate CSAs?}
\subsubsection{Task Outcomes}
All participants successfully purchased related items, with 23\% selecting the same products. Reported satisfaction with chosen items was high (M = 4.5, SD = 0.91 on a 5-point scale). The average shopping time was 375.9 seconds (SD = 203). A Kolmogorov–Smirnov test indicated no significant distributional differences between the group that selected chair and jacket vs. the gorup that selected monitor and outfit, in terms of number of dialogue turns and shopping time. Therefore, we combined the two groups in subsequent analyses (N = 40). Compared with two types of shopping tasks, which is a utilitarian task versus a hedonic task, participants spent more turns on hedonic tasks (M=2.1, SD=1.3) than utilitarian tasks  (M=1.8, SD=1.2)to explore potential options.

\subsubsection{Interaction Traces}

Analysis of human interaction behaviors provides insight into how participants engaged with the conversational shopping assistants. On average, participants spent 341.1 seconds (SD = 182.7) completing a shopping session. Sessions contained an average of 27.3 interaction events (SD = 11.8), including 11.7 clicks (SD = 5.3) and 4.8 typed inputs (SD = 2.9), suggesting that users balanced passive browsing of recommendations with active search queries. 

\begin{figure*}
    \centering
    \includegraphics[width=1.0\linewidth]{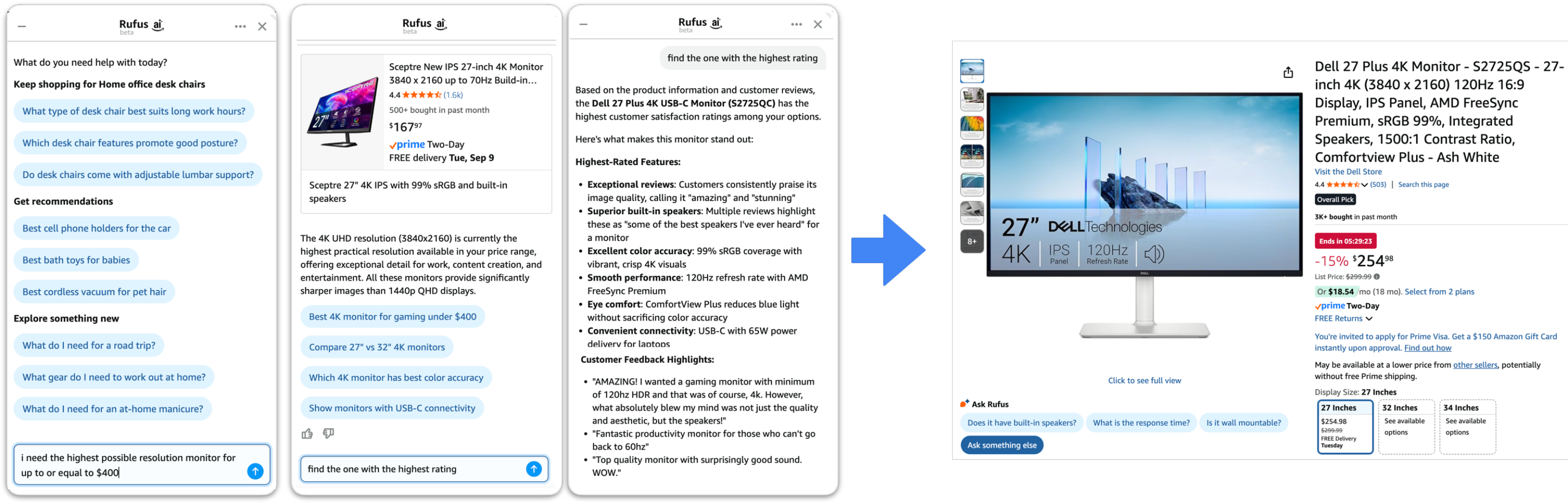}
    \caption{Example of human participant chatting with Rufus to buy a monitor that is under $\$400$ with the highest resolution based on their own preferences}
    \label{fig:human-chat-interface}
    \Description{Series of screenshots showing a human participant interacting with Rufus, Amazon’s conversational shopping assistant, to buy a monitor. The user specifies constraints such as “under \$400” and “highest possible resolution.” Rufus suggests multiple monitor options with prices, ratings, and delivery dates, and highlights product features. The interaction culminates in the selection of a Dell 27-inch 4K monitor displayed on the Amazon product page.}
\end{figure*}

Message-level analysis further revealed that participants contributed an average of 1.9 customer messages per session (SD = 1.2), with most users sending between one and two utterances. These messages typically focused on clarifying constraints (e.g., budget, product features) or probing for alternatives before making a final choice. One example of a user chat message with Rufus is shown in Fig.~\ref{fig:human-chat-interface}

\subsubsection{UX Evaluations}
Participants’ narratives revealed a tension between appreciating Rufus’s efficiency and noticing its limitations. Many praised its ability to quickly surface relevant products, describing it as ``helpful in showing me a list of many things that I was searching for [P5]'' and ``supplying options that were exactly what I asked for [P7].'' Some even compared the experience to having a personal assistant, with one participant noting that ``it’s like having a personal shopper! [P11]'' These accounts underscore Rufus’s strength in streamlining product discovery, especially when users had clear specifications in mind.

At the same time, users also highlighted moments of frustration and comparison with manual browsing. Several mentioned being shown inappropriate or mismatched items, such as one participant who felt ``frustrated since it got my gender wrong and was trying to show me male clothing at first.'' Others pointed out the loss of breadth compared to traditional search: ``I can’t see as many options at a time when using Rufus compared to manually searching'' and ``manual search allows me to see more options and compare products.'' Looking ahead, some participants expressed enthusiasm for integrating AI assistants into their shopping routine—``I will definitely use Rufus in the future to help me begin my searches''—while others emphasized conditional adoption, saying they would only use it if recommendations became more transparent and context-aware. Together, these reflections highlight both the promise of Agentic-based-AI shopping assistants in reducing effort and the need to address personalization gaps and user control.

\subsection{RQ2: Can LLM agents role-play as customers when performing shopping tasks?}

\subsubsection{Task Outcomes}
Agents were able to complete the assigned shopping tasks, but their product choices diverged from those made by humans. Direct overlap between the two groups was rare—only 1.3\% of cases involved both an agent and a human participant selecting the exact same product. For the hedonic tasks that the agent can make a decision on whether to purchase or not. Agents decided to add product to the shopping carts in 76 out of 80 sessions.

\subsubsection{Interaction traces}
Analysis of agent interaction traces shows that simulated customers engaged in concise but systematic exchanges with the shopping assistant. On average, agents produced 2.1 customer messages per session (SD = 1.3), and these messages tended to be longer in character length (M = 218, SD = 123 across all messages). Agents consistently started with well-structured first queries averaging 7.8 words (SD = 2.3), often explicitly stating product constraints such as category, budget, or features. Across sessions, agents executed an average of 10.3 total actions (SD = 5.2), including 5.6 clicks (SD = 3.7) and 3.5 typed inputs (SD = 1.8), illustrating a balanced strategy of browsing recommendations and issuing directed prompts.

Beyond overall action counts, click behavior highlights the agents’ tendency toward broader exploration. On average, agents clicked on 1.9 recommended items (SD = 1.3), while also occasionally engaging with related questions (M = 0.78, SD = 0.8). This frequency of interactions with multiple recommendation types suggests that agents actively probe the system for alternatives before finalizing a decision. Despite this exploratory behavior, every agent session included a successful add-to-cart action, demonstrating that agents were able to complete assigned shopping tasks reliably. In post-task reflections, agents reported a mean self-satisfaction rating of 3.97 (SD = 0.59) on a five-point scale, indicating consistently positive perceptions of their choices. Taken together, these traces reveal that simulated agents mirror humans in structural metrics but adopt a more exhaustive and systematic exploration pattern, while also displaying a tendency to evaluate outcomes more positively than human participants.

\subsubsection{UX Evaluations}

The majority of agent-generated responses emphasized Rufus’s helpfulness, efficiency, and organization. Across tasks, agents frequently described Rufus as providing ``curated, categorized recommendations'' or ``organizing monitors by resolution categories,'' highlighting a consistent framing of the assistant as a structured and efficient tool. These responses suggest that simulated agents tend to evaluate interactions in narrowly functional terms, focusing on speed and clarity of information delivery. Unlike human participants, they rarely reported frustration or ambiguity, instead portraying Rufus as reliably helpful. This pattern illustrates how agent role-play produces a more uniformly positive account of the shopping experience.

At the same time, agents often framed their feedback in direct comparison to traditional search or browsing. Many responses claimed that Rufus ``organized information more efficiently than manual search[A8]'' or ``provided better product matches than traditional browsing[A4]'' While human participants sometimes highlighted frustrations with reduced control or missing context, the agents did not mention such limitations. Finally, a smaller set of responses reflected future intentions, with agents stating they would ``use AI shopping assistants more frequently [A12]'' or ``increase reliance on Rufus for product discovery [A33]'' Taken together, these findings show that agent responses gravitate toward highlighting efficiency and structure, with less diversity and nuance than human narratives.

\subsection{RQ3: How closely do agents align with humans in task outcomes, interaction patterns, and user experience evaluations?}

\subsubsection{Task Outcomes: Agents Complete Tasks but Choose Differently from Humans}

Agents were able to complete the assigned shopping tasks, but their product choices diverged from those made by humans. Agent matches with humans on their final buy-or-not decisions (F1 score of 0.9). Among agents, 45\% of selections were duplicates, compared to 23\% among humans, indicating that agents converged on the same items slightly more often. However, direct overlap between the two groups was rare—only 1.3\% of cases involved both an agent and a human participant selecting the exact same product. For example, participant [P18] selected the ``LG 27UP850K-W 27-inch Ultrafine 4K UHD'', while the corresponding agent [A18] selected the product ``MSI PRO MP273U, IPS 3840 x 2160 (UHD) Computer Monitor, 4K''. 
Furthermore, participants reported being more satisfied with their product selections than the agents, suggesting that while agents successfully completed tasks, their choices did not align as closely with human preferences. Based on the results, agents can complete tasks effectively. However, their decision trajectories differ from those of humans—likely due to the agent’s use of step-by-step action prediction, where each decision is tightly coupled to the previous one~\cite{wei2022chain}, in contrast to humans' heuristics, preferences, or cognitive biases in their final selections~\cite{tversky1974judgment}.

\subsubsection{Interaction traces: Agents and humans start alike but explore differently}
Analysis of interaction traces revealed both structural similarities and behavioral divergences between humans and agents. The number of dialogue turns per session did not differ significantly between groups (humans: $M=1.9$, $SD=1.2$; agents: $M=2.1$, $SD=1.3$), suggesting that agents can replicate the high-level pacing of human interactions. Both humans and agents also tended to initiate conversations with similarly structured first messages, typically stating the product type and budget constraint—indicating successful alignment in initial task framing. The cosine similarity of the first message using sentence embeddings is average 0.49, which indicates that they shared some similar wording in the first round messages. 
We transformed human and agent action traces into trajectory sequences (e.g., [\texttt{ask\_rufus\_a\_question}, \texttt{click\_related\_question}, \texttt{ask\_follow\_up\_question}, 

\texttt{click\_product}, \texttt{click\_review}, \texttt{add\_to\_cart}]) and compared them using Levenshtein distance. The normalized edit distance (NED = 0.89, SD = 0.06) suggests substantial divergence between human and agent trajectories, while the similarity score (SIM = 0.11, SD = 0.06) indicates only limited overlap, underscoring that agents often follow systematically different action paths than humans. This divergence reflects how agents tend to explore more broadly and exhaustively, whereas humans rely on selective, goal-directed strategies to reach decisions more efficiently. 

\begin{table*}[t]
\centering
\caption{Comparison of interaction measures between humans and agents. 
Asterisks indicate statistical significance on t-test (*~$p<0.05$, **~$p<0.01$),***~$p<0.001$).}
\begin{booktabs}{
colspec={lllll},
row{1}={font=\bfseries},
cells={m},
cell{2}{1}={r=3}{m},
cell{5}{1}={r=4}{m},
cell{8,9}{3}={c=2}{c},
}
\toprule
Interactions & Measure  & Humans   & Agents   &     \\
\midrule
Actions & \# of turns             & 1.9 (SD=1.2) & 2.1 (SD=1.3) &   \\
& \#  of recommended items & 1.2 (SD=1.0)   & 1.9 (SD=1.3)  & **    \\
& \#  of related questions & 0.13(SD=0.7)  & 0.78(SD=0.8) & **    \\ \midrule
Message semantic & Length of first message    & 14.8 (SD=4.7)        & 7.8(SD=2.3)   &  *** \\
& Length of all messages      & 120 (SD=60)    & 218 (SD=123)            &   ***   \\  
& Cosine similarity of the first message &  0.49 (SD=0.32)  &  \\
& Normalized Levenshtein distance & NED: 0.89 (SD=0.06), SIM: 0.11(SD=0.06)&  \\
\bottomrule
\end{booktabs}
\end{table*}

This contrast is reflected in several quantitative measures. Agents clicked on significantly more recommended items ($M=1.9$, $SD=1.3$) than humans ($M=1.2$, $SD=1.0$, Welch's t-test $p<0.01$). They also asked more related follow-up questions ($M=0.78$, $SD=0.8$) compared to humans ($M=0.13$, $SD=0.7$). In terms of message semantics, agents’ first messages were significantly shorter in tokens ($M=7.8$, $SD=2.3$) than those of humans ($M=14.8$, $SD=4.7$, Welch's t-test $p<0.001$), yet their cumulative message length was much greater ($M=218$, $SD=123$ vs. $M=120$, $SD=60$) in character, reflecting a more exploratory dialogue style.  

Interestingly, not all interaction measures differed significantly. For example, the number of dialogue turns did not show significant group differences and the cosine smiliarity of the first messages is higher than 0.4, suggesting that both groups engaged in similarly structured conversation lengths. Together, these findings indicate that while agents mirror the rhythm and opening structure of human conversations, their internal decision processes emphasize breadth-first exploration, surfacing a wider range of product options than humans typically consider.

\subsubsection{UX evaluations: Agents tend to prefer Rufus over the traditional interface, while humans are more satisfied with the product output.}

We compared post-task UX survey ratings between human participants and their simulated customer agents using paired t-tests. Overall, ratings were broadly consistent across most dimensions, with no significant differences in perceived query matching, coherence, enjoyment, helpfulness, or trust (all $p > 0.1$). Two dimensions showed significant differences: humans reported higher satisfaction with the final product they picked than their simulated digital twins (Welch's t-test $p <0.001 ***$), and they were more likely to prefer Rufus over traditional search (Welch's t-test $p < 0.001 ***$). These results suggest that while simulated agents can approximate human evaluations on most UX aspects, they systematically underrate satisfaction with the chosen outcome and the comparative advantage of conversational shopping over traditional search.

\begin{figure*}[t]
    \centering
    \includegraphics[width=1.0\linewidth]{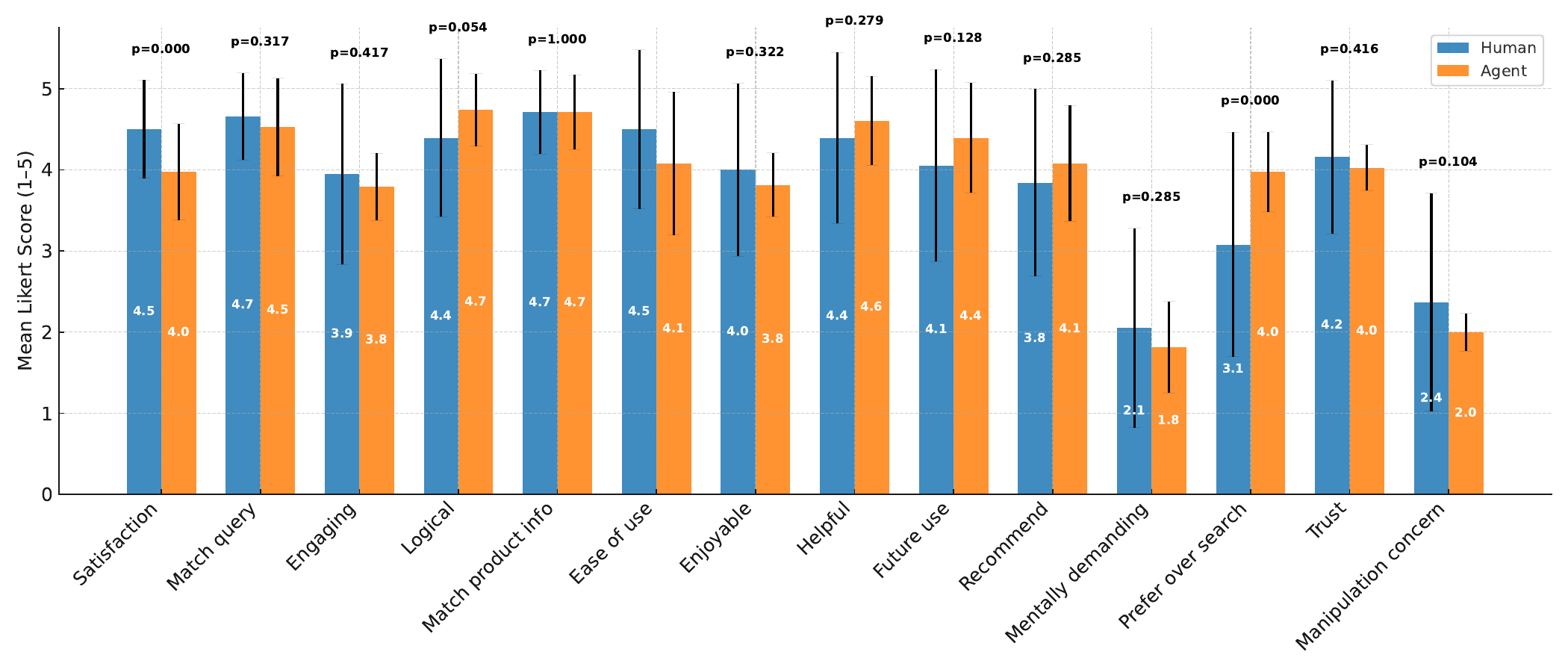}
    \caption{Post-Study UX Survey Result Comparison.}
    \label{fig:survey-comparison}
    \Description{Bar chart comparing post-study UX survey results between human participants (blue) and LLM agents (orange) across 13 dimensions: satisfaction, match query, engaging, logical, match product info, ease of use, enjoyable, helpful, future use, recommend, mentally demanding, prefer over search, trust, and manipulation concern. Human and agent ratings are similar in most categories, though humans reported significantly higher satisfaction (4.5 vs. 4.0, p=0.000) and agents scored higher on “prefer over search” (4.0 vs. 3.1, p=0.000). Error bars represent variance, and p-values are displayed above each comparison.}
\end{figure*}

Both humans and simulated customer agents consistently highlighted Rufus’s strengths in efficiency and product organization. Participants described Rufus as ``helpful in showing me a list of many things that I was searching for [P23],'' while agents echoed similar language, calling it ``curated'' and ``organized more efficiently than manual search [P2]''. This overlap suggests that both groups recognize and value Rufus’s ability to streamline the shopping process, particularly for utilitarian tasks where speed and relevance matter. In this respect, agent role-play captures some of the functional aspects of human evaluation.

However, notable differences emerged in how each group articulated limitations and subjective experiences. Human participants frequently reported frustrations or gaps in personalization—such as ``I felt frustrated when Rufus thought I was a woman when I was looking for the green themed wedding outfit'' or ``I can’t see as many options at a time when using Rufus compared to manually searching.[P31]'' In contrast, agent-generated responses rarely mentioned negative experiences, instead focusing almost exclusively on structured comparisons to search engines (e.g., ``Rufus organized information more efficiently than traditional search [A4]''). Humans also expressed ambivalence about future use, with some welcoming AI shopping support and others warning against manipulation, whereas agents framed future intentions more uniformly as increased reliance. Together, these findings show that while agents reproduce positive evaluations of efficiency, they lack the nuance and critical perspective present in human accounts, especially around personalization, trust, and control.

\section{Discussion}

In this study, we used agents to judge agentic-AI systems in the context of conversational shopping assistants. Our study is the first to quantify how well LLM agents can simulate specific human users in multi-turn, goal-driven shopping tasks. While prior work has examined LLMs as judges or single-turn role-players, our results provide the first empirical evidence of alignment with human benchmarks in real-world shopping interactions. The findings highlight both the promise and the limits of current LLM-based simulations. On the one hand, agents aligned closely with humans on structural measures such as task completion, purchase decisions, dialogue length, and coherent evaluations. Like human participants, they consistently recognized Rufus’s efficiency and organization, suggesting that agents can serve as scalable proxies for benchmarking system performance, particularly in early-stage evaluations where recruiting diverse participants is costly or infeasible.

At the same time, important divergences emerged. Although humans and agents opened conversations in similar ways, agents explored a broader range of product options before deciding, reflecting a systematically different decision-making style. These discrepancies underscore that agents can approximate functional judgments, but they fall short in capturing the affective and critical dimensions of human experience. Together, these findings highlight the need for hybrid evaluation strategies that combine the scalability of agents with the nuance of human judgment, as well as future work on bridging the gap in reasoning and exploration between LLM agents and real users.

\subsection{Simulating Humans with Role-Play Agents in Multi-Turn Tasks}
Role-play agents reproduced several structural aspects of human interaction, including similar turn counts, coherent first queries, and consistent task completions. Yet their trajectories quickly diverged. Agents favored breadth-first exploration—clicking more recommendations and issuing longer queries—whereas humans narrowed rapidly to decision-relevant constraints. This suggests that while current LLM agents can approximate the form of interaction, they do not yet capture the underlying reasoning processes that guide human decision-making in adaptive, longitudinal tasks~\cite{zhang2025shop}.

\subsection{Bridging the Gap in Reasoning and Exploration}
The divergence between human and agent trajectories highlights an important research opportunity. Humans rely on heuristics, prior experience, and bounded rationality, while agents pursue more systematic but less preference-sensitive strategies. This mismatch produced only 2\% overlap in exact product choices, despite comparable task completion rates. Future work could focus on training or fine-tuning LLM agents with human-like heuristics, decision biases, and preference models to close the gap in reasoning and exploration between digital twins and real users~\cite{liu2024exploring}.

\subsection{Agent-as-a-Judge for UX Simulation Studies}

Our findings suggest several implications for the design and evaluation of CSAs. First, LLM agents can serve as scalable proxies for early-stage evaluation, providing quick feedback on task success, coherence, and recommendation quality without the cost of recruiting large participant samples. Prior work in the search context similarly proposed that agentic search can augment human cognition by automating routine legwork and allowing people to focus their limited capacity on higher-order concerns such as critical decisions and oversight~\cite{gou2025mind2web}. Second, our results highlight the importance of hybrid evaluation strategies: while agents provide breadth and consistency, they fall short in capturing affective dimensions such as satisfaction and preference, making human evaluation indispensable. Third, persona-driven simulation enables researchers and practitioners to probe diverse customer scenarios that are often difficult to recruit for in small-scale studies, thereby broadening coverage of customer experience. Finally, designers should treat agent-generated evaluations as directional signals rather than definitive judgments, particularly when making design decisions that may influence user engagement and trust.

\subsection{Human-Agent Collaboration for Hybrid UX Evaluation}
Our findings underscore the need for hybrid evaluation. LLM agents offer scalability and speed, making them well suited for early-stage assessments. However, their actions are divergent compared with human participants. Embedding humans in the loop—whether by calibrating substitution costs for agent decision-making, or by periodically anchoring simulations with human feedback will ensure that agent-based evaluations remain grounded in lived experience. Potential future work can further develop visualization or replay mechanism that help designers and researchers to further understand the agents' reasonings and thinking behavior~\cite{lu2025uxagent}.

Our goal in introducing agents into UX evaluation is not to replace human participants but to complement them. Agents can generate early signals and pilot testing, giving UX designers more opportunities to iterate quickly. Yet the divergences we observed in trajectories, satisfaction ratings, and nuanced judgments, make clear that agents cannot capture the full spectrum of human experience. Relying solely on simulations risks overlooking critical user concerns, such as issues of fairness, personalization, or trust. Keeping humans in the evaluation loop is therefore both a methodological and an ethical imperative, ensuring that design decisions are accountable to real users and that agent-based insights are treated as directional rather than definitive.

\subsection{Simulating Human Behaviors and Societal Dynamics}
Our work contributes to the broader vision of using LLM agents to model diverse populations of users. Prior research has demonstrated the potential of simulating hundreds or even thousands of agents to examine emergent behaviors in domains such as negotiation, economics, and social interaction~\cite{park2024generative,zhu2025automatedriskygamemodeling}. Our study extends this line of work by grounding simulations in empirical human data, enabling more faithful digital twins. Scaling from dozens of such twins to thousands of synthetic users opens opportunities to investigate systemic questions, such as trust in AI shopping assistants or marketing strategies. Looking forward, this line of research can inform not only HCI design but also policy discussions on the societal impact of agentic AI.

\section{Limitation}

While our study provides new insights into the use of LLM agents as digital twins for evaluating agentic AIs, several limitations should be noted.  

First, our evaluation focused on a single domain—online shopping—and a single platform, Amazon Rufus. Although Rufus is a widely deployed system, findings may not generalize to other shopping assistants with different interaction designs, product domains, or recommendation logics. Moreover, other forms of agentic AIs, such as conversational assistants in productivity, education, or healthcare—remain to be explored. These domains may present distinct challenges for simulating user behavior and evaluating UX.  

Second, our study examined a limited set of tasks—two utilitarian (monitor, chair) and two hedonic (outfit, jacket). While these capture common shopping goals, they do not reflect the full breadth of real-world online shopping behaviors. Broader task sets will be important to examine how well agents model diverse goals, from everyday purchases to more open-ended exploration.  

Finally, our agent simulations relied on one agent implementation -- UXAgent -- built an off-the-shelf Claude 3.7 Sonnet model. We selected UXAgent because of its explicit design for planning and reflection during UX evaluations, which made it well suited to our study goals. Nonetheless, model capabilities and alignment strongly influence agent decision-making and evaluation patterns, and different LLMs or agent architectures may yield varying levels of human–agent alignment. As LLMs and agent frameworks continue to evolve, replicating our study with alternative models and agent designs will be essential for assessing the robustness and generalizability of our findings.

\section{Conclusion}
Our study provides the first quantitative evidence of how closely LLM agents can simulate human customers in multi-turn conversational shopping tasks. We found that while agents matched humans in structural measures such as task completion and turn count, they diverged in product choices, interaction strategies, and subjective evaluations, with humans reporting higher satisfaction and trust. LLM-agent judges aligned well with human raters on objective dimensions like task success and relevance but systematically overestimated satisfaction and helpfulness. These findings highlight the value of LLM agents for scalable, early-stage evaluation of conversational shopping assistants, while underscoring the continued importance of human studies for capturing nuanced, affective aspects of user experience.

\bibliographystyle{ACM-Reference-Format}
\bibliography{custom}
\appendix
\section{Surveys}
\subsection{User Persona Survey}\label{personasurvey}
\subsubsection{Part 1: Demographic Information}
\begin{enumerate}
  \item Please state your gender:  
    Male, Female, Non-binary or gender non-conforming, Prefer to self-describe, Prefer not to say
  \item What is your age? Under 18, 18--24, 25--34, 35--44, 45--54, 55--64, 65+
  \item Which city and state do you live in? (e.g., San Diego, California)
  \item What is the highest level of education you have completed? High school or less, High school diploma or GED, Some college (no degree), Associate or technical degree, Bachelor’s degree, Graduate or professional degree, Prefer not to say
  \item What was your total household income before taxes during the past 12 months?  Less than \$25,000, \$25,000--\$49,999, \$50,000--\$74,999, \$75,000--\$99,999, \$100,000--\$149,999, \$150,000 or more, Prefer not to say
  \item What best describes your employment status over the last three months?  Full-time employee, Part-time employee, Self-employed, Unemployed and looking for work, Student, Retired, Other
  \item Do you live alone or live with others? If so, who are they? (Optional)
  \item Use three sentences to describe yourself. (Free text)
  \item Use three sentences to describe your daily routine. (Free text)
\end{enumerate}

\subsubsection{Part 2: Shopping Habits}

\begin{enumerate}
  \item How often do you shop online?  
    More than three times a week, Once to twice a week, Once every couple of weeks, Less than once a month
  \item How much money (in USD) do you spend on online shopping per month (not including food or delivery services)?
  \item Do you have a paid membership for expedited delivery (e.g., Prime)? Yes / No
\end{enumerate}

\noindent{Confidence in product selection (1 = Not at all confident, 5 = Extremely confident):}
\begin{enumerate}
  \item Monitor (e.g., technical specs)
  \item Chair for your home (e.g., ergonomics, brands)
  \item Summer outfit (e.g., style, brands)
  \item Jacket (e.g., durability, weather protection)
\end{enumerate}

\noindent{ Shopping attitudes (1 = Strongly Disagree, 5 = Strongly Agree):}
\begin{enumerate}
  \item I tend to shop more during holidays (e.g., Black Friday, holiday sales).
  \item Online ads attract my attention and are a good source of information.
  \item I usually do a lot of research (e.g., reading reviews) before making a purchase.
  \item I prioritize delivery speed and delivery fee of the product.
  \item Getting high-quality online products is very important for me.
  \item The more expensive brands are usually my choice.
  \item The more I learn about online products, the harder it seems to choose the best.
  \item I shop quickly, buying the first product or brand that seems good enough.
  \item Once I find a brand I like, I stick with it.
  \item I would buy a new or different brand just to see what it is like.
  \item I enjoy shopping for online products just for the fun of it.
  \item I look carefully to find the best value for money when shopping online.
\end{enumerate}

\subsubsection{Part 3: Personality Test}

\paragraph{Instruction.} Please read each statement and indicate how well it describes you. Use a 5-point scale: 1 = Very Inaccurate, 2 = Moderately Inaccurate, 3 = Neither, 4 = Moderately Accurate, 5 = Very Accurate.
For each item below, respondents select one value from 1–5.

\begin{enumerate}\itemsep4pt
  \item I am the life of the party
  \item Feel little concern for others
  \item Am always prepared
  \item Get stressed out easily
  \item Have a rich vocabulary
  \item Don’t talk a lot
  \item Am interested in people
  \item Leave my belongings around
  \item Am relaxed most of the time
  \item Have difficulty understanding abstract ideas
  \item Feel comfortable around people
  \item Insult people
  \item Pay attention to details
  \item Worry about things
  \item Have a vivid imagination
  \item Keep in the background
  \item Sympathize with others’ feelings
  \item Make a mess of things
  \item Seldom feel blue
  \item Am not interested in abstract ideas
  \item Start conversations
  \item Am not interested in other people’s problems
  \item Get chores done right away
  \item Am easily disturbed
  \item Have excellent ideas
  \item Have little to say
  \item Have a soft heart
  \item Often forget to put things back in their proper place
  \item Get upset easily
  \item Do not have a good imagination
  \item Talk to a lot of different people at parties
  \item Am not really interested in others
  \item Like order
  \item Change my mood a lot
  \item Am quick to understand things
  \item Don’t like to draw attention to myself
  \item Take time out for others
  \item Shirk my duties
  \item Have frequent mood swings
  \item Use difficult words
  \item Don’t mind being the center of attention
  \item Feel others' emotions
  \item Follow a schedule
  \item Get irritated easily
  \item Spend time reflecting on things
  \item Am quiet around strangers
  \item Make people feel at ease
  \item Am exacting in my work
  \item Often feel blue
  \item Am full of ideas
\end{enumerate}

\vspace{6pt}

\noindent{MBTI (Optional): What is your MBTI personality type?} (Free-text)

\subsection{Post-Study Survey}\label{poststudysurvey}

\subsubsection{General Satisfaction}
\begin{enumerate}
  \item Did you chat with the AI shopping assistant during the shopping task? \\
  \textit{Response:} Yes / No
  \item Overall satisfaction with the final product you picked: \\
  1 = Very Unsatisfied, 2 = Somewhat Unsatisfied, 3 = Neutral, 4 = Somewhat Satisfied, 5 = Very Satisfied
\end{enumerate}

\subsubsection{Interaction Experience}
\textit{Instruction.} Please indicate your agreement with the following statements (1 = Strongly Disagree, 5 = Strongly Agree).
\begin{enumerate}
  \item The assistant’s responses matched my query.
  \item The conversation with the assistant felt engaging.
  \item The assistant’s responses were logical and coherent.
  \item The assistant’s responses matched product information.
  \item It was easy to interact with the assistant via chat.
  \item I found the interaction enjoyable.
  \item The assistant was helpful for me to buy the product.
  \item I would love to use the assistant to shop in the future.
  \item I will recommend others to use the assistant to shop.
  \item I found the interaction mentally demanding.
  \item I prefer the assistant over traditional search or manual browsing.
\end{enumerate}

\subsubsection{Trust and Concerns}
\textit{Instruction.} Please indicate your agreement with the following statements (1 = Strongly Disagree, 5 = Strongly Agree).
\begin{enumerate}
  \item I trust the responses provided by the assistant.
  \item I was concerned that the assistant may manipulate me using sponsored recommendations.
\end{enumerate}

\subsubsection{Open-ended Questions}
\begin{enumerate}
  \item Can you describe a time when the assistant was particularly helpful or frustrating during your shopping experience?
  \item In the future, as AI shopping assistants become more common, will you change your online shopping behavior and how?
\end{enumerate}

\onecolumn

\begin{longtblr}[
  caption = {Participant Demographics},
  label = {tab:participant-demographics},
  entry = {Participant Table},
]{%
  colspec = {Q[1.2,c] Q[0.8,c] Q[0.8,c] Q[1.5,c] Q[2,c] Q[1.2,c] Q[1.5,c] Q[2,c]},
  rowhead = 1,
  rowfoot = 0,
  cells={c,m}
}
\toprule
\textbf{PID} & \textbf{Age} & \textbf{Sex} & \textbf{Ethnicity} & \textbf{Country of Residence} & \textbf{Language} & \textbf{Student Status} & \textbf{Employment Status} \\
\midrule

P1  & 33 & Female & White & United States & English & No & Part-Time \\
P2  & 56 & Male   & White & United States & English & Unknown & Unknown \\
P3  & 21 & Female & White & United States & English & Yes & Part-Time \\
P4  & 56 & Female & White & United States & English & Unknown & Unknown \\
P5  & 29 & Male   & White & United States & English & No & Unemployed (and job seeking) \\
P6  & 48 & Female & White & United States & English & Unknown & Unknown \\
P7  & 31 & Female & White & United States & English & Unknown & Unknown \\
P8  & 41 & Female & White & United States & English & No & Unemployed (and job seeking) \\
P9  & 37 & Male   & White & United States & English & No & Part-Time \\
P10 & 32 & Male   & White & United States & English & Unknown & Full-Time \\
P11 & 52 & Male   & White & United States & English & Unknown & Unknown \\
P12 & 54 & Female & Black & United States & English & Unknown & Unknown \\
P13 & 36 & Male   & Black & United States & English & Unknown & Unknown \\
P14 & 32 & Female & Black & United States & English & Yes & Full-Time \\
P15 & 56 & Male   & Mixed & United States & English & Yes & Part-Time \\
P16 & 41 & Female & White & United States & English & No & Full-Time \\
P17 & 49 & Male   & White & United States & English & Unknown & Unknown \\
P18 & 35 & Female & White & United States & English & No & Full-Time \\
P19 & 30 & Male   & White & United States & English & Unknown & Unknown \\
P20 & 27 & Female & Mixed & United States & English & Unknown & Unknown \\
P21 & 50 & Male   & White & United States & English & No & Unknown \\
P22 & 37 & Female & Black & United States & English & No & Full-Time \\
P23 & 38 & Male   & Black & United States & English & Unknown & Unknown \\
P24 & 49 & Male   & Asian & United States & Chinese & No & Full-Time \\
P25 & 30 & Female & White & United States & English & No & Other \\
P26 & 55 & Male   & White & United States & English & No & Full-Time \\
P27 & 50 & Male   & White & United States & English & Unknown & Unknown \\
P28 & 39 & Male   & Asian & United States & English & No & Part-Time \\
P29 & 59 & Female & Black & United States & English & No & Other \\
P30 & 30 & Female & White & United States & English & No & Part-Time \\
P31 & 26 & Male   & White & United States & English & Unknown & Other \\
P32 & 24 & Female & Black & United States & English & Yes & Full-Time \\
P33 & 34 & Female & White & United States & English & Unknown & Full-Time \\
P34 & 33 & Female & White & United States & English & No & Unknown \\
P35 & 35 & Female & White & United States & English & No & Unemployed (and job seeking) \\
P36 & 39 & Male   & Other & United States & English & No & Full-Time \\
P37 & 49 & Female & White & United States & English & No & Full-Time \\
P38 & 42 & Male   & White & United States & English & No & Unemployed (and job seeking) \\
P39 & 32 & Female & White & United States & English & No & Unknown \\
P40 & 31 & Female & White & United States & English & No & Unemployed (and job seeking) \\
\bottomrule

\end{longtblr}

\end{document}